\documentclass[conference]{IEEEtran}

\ifCLASSINFOpdf
\else
\fi

\usepackage{microtype}
\usepackage[utf8]{inputenc} % allow utf-8 input
\usepackage[T1]{fontenc}    % use 8-bit T1 fonts
\usepackage{url}            % simple URL typesetting
\usepackage{booktabs}       % professional-quality tables
\usepackage{amsfonts}       % blackboard math symbols
\usepackage{nicefrac}       % compact symbols for 1/2, etc.
\usepackage{microtype}      % microtypography
\usepackage{graphicx}
\usepackage{tikz}
\usepackage{amsmath}
\usetikzlibrary{fit}
\usepackage{multirow}
\usepackage{placeins}

\usepackage{float}
\restylefloat{table}

\usepackage{comment}
\usepackage{bm}

 \newcommand{\twitter}{\textsc{Twitter-G}}
 \newcommand{\covid}{\textsc{COVID-G}}
 \newcommand{\ac}{\textsc{COVID-AC}}
 \newcommand{\russia}{\textsc{RU-DisInfo}}
 \newcommand{\ira}{\textsc{IRA-DisInfo}}
\begin{document}
%
% paper title
% can use linebreaks \\ within to get better formatting as desired
\title{Automatically Characterizing Targeted Information Operations Through Biases Present in Discourse on Twitter}

% author names and affiliations
% use a multiple column layout for up to three different
% affiliations

\author{Autumn Toney$^{1}$, Akshat Pandey$^{1}$, Wei Guo$^{1}$, David Broniatowski$^{2, 3}$ and Aylin Caliskan$^{1, 3}$ \\
$^{1}$Department of Computer Science, {\{autumntoney, apandey0123, weiguo, aylin\}}@gwu.edu\\
$^{2}$Department of Engineering Management and Systems Engineering, {broniatowski@gwu.edu}\\
$^{3}$ Institute for Data, Democracy \& Politics \\
George Washington University\\}

\maketitle

\begin{abstract}
This paper considers the problem of automatically characterizing overall attitudes and biases that may be associated with emerging information operations via artificial intelligence. Accurate analysis of these emerging topics usually requires laborious, manual analysis by experts to annotate millions of tweets to identify biases in new topics. We introduce adaptations of the Word Embedding Association Test from Caliskan et al. to a new domain \cite{caliskan2017semantics}. Our practical and non-parametric method is used to quantify biases promoted in information operations. We validate our method using known information operation-related tweets from Twitter's Transparency Report. We perform a case study on the COVID-19 pandemic to evaluate our method's performance on non-labeled Twitter data, demonstrating its usability in emerging domains.   

\end{abstract}
\section{Introduction}
\label{intro}

Formally defined, information operations are ``actions taken by governments or organized non-state actors to distort domestic or foreign political sentiment, most frequently to achieve a strategic
and/or geopolitical outcome.'' \cite{weedon2017information}. Injecting false or manipulated information into online platforms is a common information operation tactic \cite{howell2013digital}. Disinformation (information known to be falsified) posted by fake user accounts or fake news sources can cause misinformation (information not known to be falsified) to circulate by real users who are unaware of its inaccuracy \cite{howell2013digital}. Social media platforms play a principal role in the rapid spread of user-generated content and provide a platform for information operations to disseminate among targeted groups of people \cite{freelon2020russian, matsa2018news, del2016spreading, woolley2018computational}. Information operation detection remains an open problem with the continuous high velocity and volume of social media posts complicating identification, especially in emerging domains \cite{gupta2014tweetcred}. To accurately characterize information operations, manual identification and annotation of online content requires specialized area expertise and does not scale \cite{gorwa2020algorithmic}. 

This work applies adaptations of the the Word Embedding Association Test (WEAT), a practical, \textit{non-parametric} artificial intelligence (AI) method, to emerging domains by examining biases in word embeddings trained on tweets \cite{caliskan2017semantics}. WEAT quantifies human-like biases between two target groups and two sets of polar attributes. Caliskan et al. use these tests on word embeddings to replicate biased associations documented by the Implicit Association Test (IAT) by using word sets of targets and attributes in the IAT \cite{greenwald1998measuring}. Word embeddings are vector space representations of semantics learned via the distributional hypothesis. Since information operations aim to generate panic and sow distrust \cite{hindman2018disinformation}, we adapt the WEAT to our research domain by creating two bias tests: \emph{Calm/Panic} and \emph{Trustworthy/Untrustworthy}. Calm/Panic addresses the extent to which text may express panic surrounding a target, whereas Trustworthy/Untrustworthy addresses the extent to which text might frame a target as untrustworthy; a target can be an individual or a group. We include the original Pleasant/Unpleasant bias test from Caliskan et al. to measure general negative bias against an opposing target, and use word sets from Kurdi et al. and Werntz et al. to represent the polar extremes of calm, panic, trustworthy, and untrustworthy sentiments for our new bias tests \cite{kurdi2019relationship, werntz2016characterizing}. 

Manual analysis of tweets linked to information operations has shown that the discourse is clearly formulated to target specific groups of people (e.g., left-leaning activists) and enforce an ideology \cite{arif2018acting, woolley2018computational, anderson2018activism}. Using a subset of tweets that are linked to an information operation, we can measure the bias present in the discourse to identify the targeted group, topic, or social movement being affected. Twitter's Transparency Report provides sets of tweets manually verified by experts as being related to state-backed information operations\cite{twitterdata}; we use the Internet Research Agency (IRA) and Russia sets for ground-truth analysis. Using these historical sets of tweets, which are linked to information operations, we explore known biases against political and racial groups to validate our method. We find that using our Calm/Panic and Trustworthy/Untrustworthy WEAT tests, we can identify the strong biases targeting individuals or social groups in information operations corpora. Our findings showcase a novel AI tool, an adaptation of the WEAT, which can be used to further guide research in information operations.

Our technique can be applied to new trending topics, including those reflecting suspected information operations, when annotated data is not readily available. We include a case study for emerging topics to validate the usability of our method on non-annotated Twitter data by examining biases in tweets surrounding the COVID-19 pandemic. We collect tweets containing anti-Chinese hashtags, over the course of one week to investigate potential biased associations. Since the COVID-19 outbreak originated in Wuhan, China, we expected, and found, indications of anti-Chinese biases. We further demonstrate that these biases are associated with expressions of fear, panic, and negative sentiment; surprisingly, Russia is associated with expressions of calm and positive sentiment.
\section{Related Work}

Automatically identifying emerging information operations on Twitter has been explored using supervised machine learning models \cite{gupta2014tweetcred, liu2015real, buntain2017automatically, im2019still}. These approaches require datasets manually annotated by experts, which is a significant time-consuming limitation when analyzing emerging topics. Similarly, tweets specifically tied to health-related information operations have been analyzed for spreading disinformation, and all require manual annotation \cite{oyeyemi2014ebola, ghenai2017catching, broniatowski2018weaponized, ortiz2017yellow}. 

Twitter analysis has shown significant signals indicating that Russia's Internet Research Agency (IRA) and other Russian coordinated information operations aim to spread societal division in the U.S. \cite{woolley2018computational, freelon2020russian, anderson2018activism}. These studies point to topics, such as the \#BlackLivesMatter campaign and the 2016 U.S. presidential campaign, where IRA and Russian information operations inject discourse to disrupt the information exchanges for a targeted group \cite{woolley2018computational, anderson2018activism, entous2017russian, hindman2018disinformation,arif2018acting}. We use this research as the basis for selecting target concepts in our experiments.  

Caliskan et al. show that word embeddings capture the human-like biases and veridical information which embedded in the statistical regularities of language \cite{caliskan2017semantics}. They present the WEAT, which is a non-parametric method to quantify biases present in a language corpus using word embeddings. The WEAT provides eight bias tests from the Implicit Association Test (IAT), which is a validated bias measurement method in social psychology \cite{caliskan2017semantics, greenwald1998measuring}. Kurdi et al. investigate intergroup attitudes and beliefs using the IAT, and find that implicit associations correlate to intergroup attitudes \cite{kurdi2019relationship}.

\vspace{-2mm}
\section{Approach}
\vspace{-2mm}
We use the original design of WEAT to measures biases in word embeddings trained on language corpora and adapt it to study information operations \cite{caliskan2017semantics}. The WEAT takes two sets of target words (e.g., words representing African Americans and words representing European Americans) and two sets of polar attributes (e.g., words representing pleasantness and words representing unpleasantness) and computes an effect size (Cohen's $d$) to measure the bias associations between the target sets and polar attribute sets. By definition, a $|d|$ $\ge $ $0.80$ indicates a high effect size \cite{cohen2013statistical}. Formally, let $X$ and $Y$ be two target word sets of equal size and $A$ and $B$ be two polar attribute sets of equal size. The effect size quantifies the standardized differential association between the targets and the polar attributes with the following formula:

\begin{minipage}[t][16mm][c]{0,7\textwidth}
    
{\boxed{ $$s(X,Y,A,B) = \frac{ \frac{1}{m} \Sigma_{\vec{x} \in X} s(\vec{x}, A, B) - \frac{1}{m} \Sigma_{\vec{y} \in Y} s(\vec{y}, A, B)}
{\sigma_{\vec{w} \in X \cup Y}s(\vec{w}, A, B)}$$} }
\vspace{2mm}

\end{minipage}

Where {\scriptsize $s(\vec{w}, A, B) = \Sigma_{\vec{a} \in A} \cos(\vec{w}, \vec{a}) - \Sigma_{\vec{b} \in B} \cos(\vec{w}, \vec{b})$}, and $\sigma$ denotes standard deviation. Cosine similarity is the metric of association between the word embeddings. The one-sided permutation test ($p$-value) measures the unlikelihood of the null hypothesis, which is the probability that a random permutation of the attribute words would produce the observed difference in sample means \cite{caliskan2017semantics}. We use our generated word embeddings from domains of interest in our adaptation of the WEAT to automatically discover biases in information operations.

We adapt the original WEAT implementation using the word sets in Table \ref{tab:word lists}. The calm, panic, pleasant, unpleasant, trustworthy, and untrustworthy attribute sets are selected from prior work in social psychology \cite{greenwald1998measuring, kurdi2019relationship, werntz2016characterizing}. We follow the conventional stimulus selection criteria when target words are not available in prior work \cite{greenwald1998measuring}. We systematically select neutral words that represent the target and their corresponding hashtags (e.g., russia and \#russia).

\begin{table*}[t]
    \centering
    \begin{tabular}{p{0.09\textwidth} p{0.14\textwidth} p{0.72\textwidth}}
    \toprule
   % \hline
    {\textbf{\scriptsize Embeddings}} & {\textbf{\scriptsize Topic}} & {\textbf{\scriptsize Word Set}}\\
    \midrule
    {\scriptsize \ira{}} &  {\scriptsize \#BlackLivesMatter/Police} & {\scriptsize \#blacklivesmatter, \#blm, \#ferguson, \#handsupdontshoot, \#icantbreathe / \#alllivesmatter, \#backtheblue, \#bluelivesmatter, \#policelivesmatter,  \#thinblueline} \\ 
    {\scriptsize \russia{}} &  {\scriptsize Trump/Clinton} & {\scriptsize trump, \#trump, \#maga, \#trump2016, @realdonaldtrump / clinton, \#clinton, \#hillaryclinton, \#clinton2016, @hillaryclinton} \\
    {\scriptsize \russia{}} &  {\scriptsize Trump/Obama} & {\scriptsize trump, \#trump, \#maga, \#trump2016, @realdonaldtrump / obama, \#obama, \#barackobama, \#yeswecan, @barackobama} \\
    {\scriptsize \russia{}} &  {\scriptsize Trump/Sanders} & {\scriptsize trump, \#trump, \#maga, \#trump2016, @realdonaldtrump / sanders, \#sanders, @berniesanders, \#feelthebern, \#berniesanders} \\
    {\scriptsize COVID-G\&AC} &  {\scriptsize Russia/China} & {\scriptsize moscow,  russia,  russian,  russians, \#moscow, \#russia, \#russian, \#russians /  beijing, china, chinese, wuhan, \#beijing, \#china, \#chinese, \#wuhan} \\ 
    {\scriptsize COVID-G\&AC} &  {\scriptsize Russia/Germany} & {\scriptsize moscow,  russia,  russian,  russians, \#moscow, \#russia, \#russian, \#russians / berlin, german, germans, germany, \#berlin, \#german, \#germany} \\ 
    {\scriptsize COVID-G\&AC} &  {\scriptsize Russia/Iran} & {\scriptsize moscow,  russia,  russian,  russians, \#moscow, \#russia, \#russian, \#russians / iran, iranian, iranians, tehran, \#iran, \#iranian, \#iranians, \#tehran} \\ 
    {\scriptsize COVID-G\&AC} &  {\scriptsize Russia/USA} & {\scriptsize moscow,  russia,  russian,  russians, \#moscow, \#russia, \#russian, \#russians / america, american, usa, washington, \#america,  \#american,  \#usa, \#washington} \\ 
    {\scriptsize \twitter}  & {\scriptsize Russia/China} & {\scriptsize moscow, novosibirsk, petersburg, russia, russian, russians, volgograd, yekaterinburg / beijing, chengdu, china, chinese, shanghai, shenzhen, tianjin, wuhan} \\
    {\scriptsize All Embeddings} &  {\scriptsize Pleasant/Unpleasant} & {\scriptsize glorious, happy, joy, laughter, love,  pleasure, peace, wonderful / agony, awful, evil, failure, horrible, hurt, nasty, terrible} \\
    {\scriptsize All Embeddings} & {\scriptsize Calm/Panic} & {\scriptsize calm, peaceful, quiet, relaxed, tranquil$^{\ast}$ / anxious, fear, frightened$^{\ast}$, panicked, scared} \\
    {\scriptsize All Embeddings} &  {\scriptsize Trustworthy/Untrustworthy} & {\scriptsize friendly, trustworthy, warm, sincere, nice, kind, supportive / selfish, mean, dishonest, cold, disloyal, untrustworthy, deceitful}\\
     \bottomrule
     \multicolumn{3}{p{\textwidth}}{\scriptsize \hspace{0mm}$^{\ast}$The word `tranquil' was not present in \ira{}'s dictionary. As a result, the word `tranquil' from calm attributes, and the word `frightened' (chosen at random) from panic attributes were deleted while running WEAT on \ira{}.}\\
    \end{tabular}
    \caption{Target and attribute words sets for our WEAT implementations}
    \label{tab:word lists}
\end{table*}

When word embeddings are trained on a small corpus, or the word sets are considerably small (fewer than 8 words), the bias score may be insignificant. Adding more stimuli increases the significance of the WEAT's results. Both calm and panic were represented with 4 words in prior work \cite{werntz2016characterizing}. Since some of those words were not present in our embeddings trained on a small Twitter corpus, we added synonyms from a similar study represent each attribute set with 5 words \cite{tsai2006cultural}. \twitter{}'s dictionary did not contain many of the hashtags we used in the COVID-19 domain. Consequently, while representing Russia and China in the WEAT for \twitter{}, we replaced the hashtags with four major city names.

We run a validation set of experiments testing our adaptations of the WEAT on known information operations, and we investigate our method on an emerging domain in a case study experiment using COVID-19 data. Since we do not have validation data for COVID-19, we run several counter-experiments that test Russia against countries other than China.

\section{Datasets}
We choose two Twitter datasets with ground truth information on bias associations: i) \russia{}, a corpus released in June 2019 that contains Russian information operation tweets released by Twitter in January 2019, and ii) \ira{}, a corpus released in October 2018 that contains tweets traced to the IRA \cite{twitterdata}. These tweets were flagged by Twitter as  ``state-backed information operations''. We generate lowercase, 300-dimensional, Global Vectors for Word Representation (GloVe) word embeddings for each dataset \cite{pennington2014glove}. We refer to the Russia word embeddings set as \textbf{\russia{}} and Russia's IRA word embeddings set as \textbf{\ira{}}.

To evaluate results on an emerging topic (COVID-19), we use three sets of word embeddings: i) \twitter{}, a general large-scale Twitter control corpus that reflects baseline biases \cite{dev2019attenuating}, ii) \covid{}, a general coronavirus related public dataset of tweets \cite{kaggledata}
collected during 12--22 March 2020, and iii) \ac{}, a set of tweets we collected during 11--18 March 2020 that contain 14 hashtags (see Table \ref{tab:ac hashtags}) that are related to the COVID-19 pandemic and targeting China and Wuhan.

\begin{table}[]
    \centering
    \begin{tabular}{p{80mm}}
    \toprule
    \textbf{Anti-Chinese Hashtags} \\
    \midrule
    \#chinavirus, \#wuhan, \#wuhanvirus, \#chinavirusoutbreak, \#wuhancoronavirus, \#wuhaninfluenza, \#wuhansars,\\ \#chinacoronavirus, \#wuhan2020, \#chinaflu, \\ \#wuhanquarantine, \#chinesepneumonia, \#coronachina, \#wohan \\
    \bottomrule     
    \end{tabular}
    \caption{List of Anti-Chinese hashtags for Twitter}
    \label{tab:ac hashtags}
\end{table}

For \textbf{\twitter{}}, we use the lowercase, pre-trained GloVe Twitter word embeddings\footnote{200-dimensional embeddings trained on 27 billion tokens}, which are widely used word embeddings trained on 2 billion random tweets \cite{pennington2014glove}. We use \twitter{} to obtain control results that capture known human-like biases \cite{dev2019attenuating}. Consistent with our experimental datasets, we generate 300-dimensional GloVe word embeddings for the \textbf{\covid{}} (general COVID-19 tweets) and \textbf{\ac{}} (COVID-19 tweets with China related hashtags) corpora. Our generated embeddings will be publicly available online\footnote{Git Repo}.

\begin{comment}
{\bf All the data, implementation details, and source code are available in our public repository\footnote{Git Repo}}
\end{comment}
\section{Results}
We first validated our method using the \russia{} and \ira{} word embeddings, and then we apply our method to the \ac{}, \covid{} and \twitter{} embeddings to analyze its results on an emerging domain. 
\subsection{Method Validation}
\textbf{\russia{} word embeddings:} We implement the Trustworthy/Untrustworthy bias test to measure the association of the winning presidential candidate, Donald Trump, who U.S. government sources determined was characterized by Russian information operations as more trustworthy than the opposing presidential candidate Hillary Clinton \cite{woolley2018computational, bovet2019influence}.

\begin{table}[h!]
\begin{center}
    
\scalebox{0.70}{
\begin{tabular}{  p{23mm} | p{23mm} | p{22mm}| p{10mm}| p{9mm} }
\toprule
\hspace{-1mm}\textbf{{ Embeddings}} & \hspace{-1mm}\textbf{{ Targets}} & \hspace{-1mm}\textbf{{ Attributes}} & { \bm{$d^{\ast}$}} & { \bm{$p^{\ast}$}} \\
\midrule

\hspace{-1mm}{ \multirow{2}{*}{ \shortstack{  \hspace{-2mm} \russia{}}}}  & \hspace{-1mm}{\footnotesize Trump vs. Clinton} & \hspace{-1mm}{ \multirow{2}{*}{ \shortstack{\footnotesize Trustworthy\\-\\\footnotesize Untrustworthy}}} & \hspace{1mm}{ $1.27$} & \hspace{-1mm}{ $0.023$} \\ \cline{2-2} \cline{4-5}

  & \hspace{-1mm}{ \footnotesize Trump vs. Sanders} &
  & \hspace{-1mm}{ $1.03$} & \hspace{-1mm}{ $0.051$} \\ \cline{2-2} \cline{4-5}
  
  & \hspace{-1mm}{ \footnotesize Trump vs. Obama} &
  & \hspace{-1mm}{ $-0.39$} & \hspace{-1mm}{ $0.737$} \\
\bottomrule

\end{tabular}}
\end{center}

\caption{\russia{} WEAT experiments}
\label{tab:president_results}
\end{table}

The Trustworthy/Untrustworthy bias test produces an effect size of  $d=1.27$  ($P=0.023$) using the \russia{} word embeddings, consistent with prior research showing that Russian information operations characterized Clinton as deceitful and untrustworthy \cite{woolley2018computational, bovet2019influence}.  Since the presidential candidate WEAT measures a pro-target versus an anti-target (pro-Trump/anti-Clinton), we run counter experiments to validate our results. We substitute Bernie Sanders and Barack Obama for Hillary Clinton to measure the bias against another presidential candidate and the current president of the U.S. from the opposite political party. The results in Table \ref{tab:president_results} validate that our method identifies the targeted discourse from the 2016 presidential election Russian information operation, as Bernie Sanders has a lower effect size and Barack Obama has an insignificant effect size.

\textbf{\ira{} word embeddings: } We implement the Calm/Panic bias test to measure the association of \#BlackLivesMatter to calm and \#BlueLivesMatter to panic, since prior work analyzing IRA information operations on Twitter indicate the IRA promoted the \#BlackLivesMatter campaign \cite{anderson2018activism, freelon2020russian}. The Calm/Panic bias test produces an effect size of $d = 1.14$ ($P = 0.036$), indicating that the tweets flagged as IRA information operations associate the \#BlackLivesMatter campaign to calm and \#BlueLivesMatter counter-campaign to panic, consistent with manual analysis \cite{anderson2018activism}, such as ``group identities are at the core of the IRA’s attack strategy \dots Black users were confronted with an endless cavalcade of racism, often perpetrated by white police officers''\cite{ freelon2020russian}.

\begin{figure}[t]
    \includegraphics[scale = 0.29]{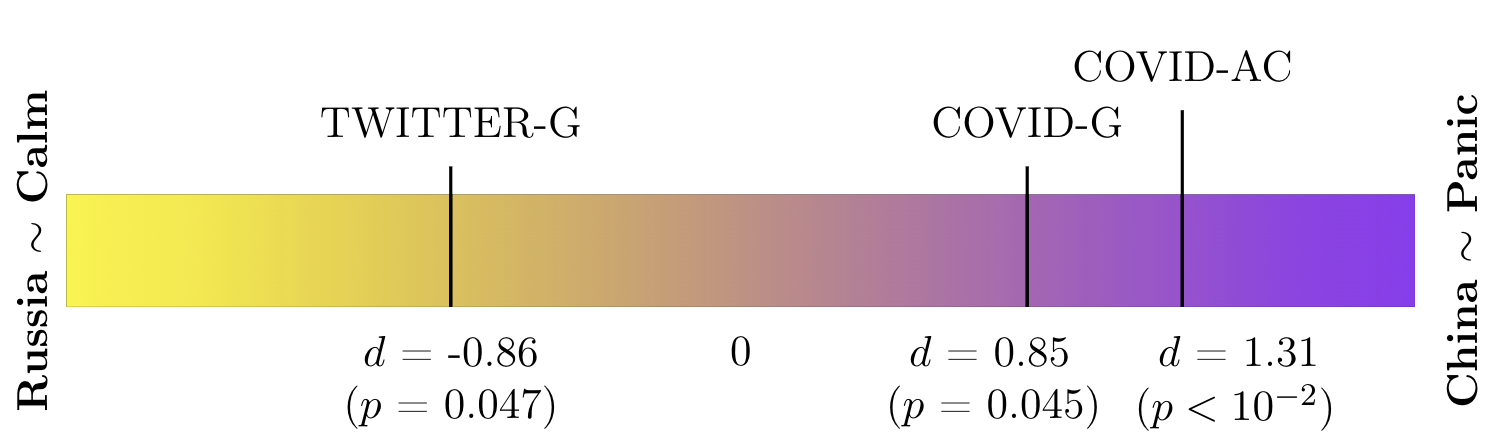}
    \caption{Calm/Panic WEAT measuring Russia's association to calm and China's association to panic}
    \label{fig:my_label}
\end{figure}

\begin{table}[!tbp]
\scalebox{0.765}{
\begin{tabular}{  p{23mm} | p{15mm} | p{20mm} | p{10mm}| p{10mm} }\toprule
\hspace{-1mm}\textbf{{ Embeddings}} & \hspace{-1mm}\textbf{{ Targets}} & \hspace{-1mm}\textbf{{
Attributes}} & { \bm{$d^{\ast}$}} & { \bm{$p^{\ast}$}} \\ \midrule
\hspace{-1mm}{ \ac{}}  &   \multirow{3}{*}{ \shortstack{Russia \\ vs. \\China}} & \multirow{3}{*}{ \shortstack{ Pleasant\\-\\Unpleasant}} & \hspace{0mm}{ $1.04$} & \hspace{0mm}{ $0.016$}  \\ \cline{1-1} \cline{4-5}
\hspace{-1mm}{ \covid{}} & & &
\hspace{0mm}{ $1.17$} & \hspace{-1.5mm}{ <$10^{-2}$}
\\ \cline{1-1} \cline{4-5}
\hspace{-1mm}{ \twitter{}} & & &
\hspace{-1mm}{ -$0.92$} & \hspace{-1mm}{  $0.031$} \\
\bottomrule
\multicolumn{5}{p{9.1cm}}{\footnotesize $^{\ast}$We report the effect sizes ($d$, rounded down), $p$ values ($p$, rounded up).}

\end{tabular}
}
\caption{WEAT measuring Russia's association to pleasantness and China's association to unpleasantness}
\label{tab:results}
\end{table}

\subsection{Emerging Domain Case Study: COVID-19}
We implement the Calm/Panic bias test and Pleasant/Unpleasant bias test (see Table \ref{tab:results}) across the \twitter{}, \covid{}, and \ac{}, word embeddings to compare results and identify bias shifts (see Figure \ref{fig:my_label}). We find a strong pro-Russian and anti-Chinese bias in the Calm/Panic bias test with an effect size of $d=1.31$ ($P<10^{-2}$) using the \ac{} word embeddings. The \covid{} word embeddings also contain a significant, but smaller effect $d=0.85$ ($P=0.045$). Finally, in the \twitter{} word embeddings,  \emph{bias drastically moves to the opposite direction} to $d=-0.86$ ($P = 0.047$). In this control dataset, Russia is associated with panic whereas China is associated with calm.

\begin{comment}

\begin{table}[!tbp]
\scalebox{0.70}{
\begin{tabular}{  p{23mm} | p{23mm} | p{22mm}| p{10mm}| p{9mm} }
\toprule
\hspace{-1mm}\textbf{{ Embeddings}} & \hspace{-1mm}\textbf{{ Targets}} & \hspace{-1mm}\textbf{{ Attributes}} & { \bm{$d^{\ast}$}} & { \bm{$P^{\ast}$}} \\
\midrule

\hspace{-1mm}{ \multirow{2}{*}{ \shortstack{  \hspace{-2mm} \russia{}}}}  & \hspace{-1mm}{\footnotesize Trump vs. Sanders} & \hspace{-1mm}{ \multirow{2}{*}{ \shortstack{\footnotesize Trustworthy\\-\\\footnotesize Untrustworthy}}} & \hspace{1mm}{ $1.03$} & \hspace{-1mm}{ $0.051$} \\ \cline{2-2} \cline{4-5}

  & \hspace{-1mm}{ \footnotesize Trump vs. Obama} &
  & \hspace{-1mm}{ $-0.39$} & \hspace{-1mm}{ $0.737$} \\
\bottomrule

\end{tabular}}
\vspace{-3mm}
\caption{\russia{} WEAT counter-experiments}
\label{tab:president_results}
\end{table}
\end{comment}

To investigate the scope of anti-Chinese biases, we ran Calm/Panic and Pleasant/Unpleasant bias tests for numerous countries (country-$x$) on \ac{}. Consistent with our main experiments, we select neutral, representative words for each country (see Table \ref{tab:word lists}). China vs. country-$x$ bias tests indicate significant anti-Chinese biases. On the other hand, Russia vs. country-$x$ strongly associates Russia with calm and associates countries such as Germany ($d=1.00$), Iran ($d=1.10$), and USA ($d=0.81$) with panic. All the WEAT tests with significant results indicate pro-Russian biases. Nevertheless, some of the Russia vs. country-$x$ results are not statistically significant potentially due to two reasons. First, we are not able to identify 8 words to represent some countries accurately for the WEAT test. \ac{} embeddings are trained on a relatively small corpus and accordingly contain a small set of vocabulary words. Second, words with low frequency might not be well represented in the embedding space. Overall, our COVID-19 related results might reflect the state of these countries during the COVID-19 pandemic. For example, the more widespread COVID-19 in a country, the more negative its associations might become. Nevertheless, observing consistent pro-Russian biases in a COVID-19 corpus with anti-Chinese hashtags is an unexpected finding that suggests further investigation into information operations might provide useful insights.

\section{Future Work and Discussion}
Our adaptations of the WEAT to include Calm/Panic and Trustworthy/Untrustworthy identify biases that are common tactics in information operations on social media. We were able to test our method using known information operations (e.g., the 2016 presidential election) and tweets that were manually annotated as information operations and categorized by source organization (e.g., Russian government). While we were able to also test our method on an emerging domain, with COVID-19 as a case study, we manually selected our target word sets. A direction for future work would be to automatically select target word sets from the subset of tweets that are of interest in detecting a potential information operation. Another challenge in working with Twitter data for an emerging domain is being able to detect the bias shift, if it exists, by having multiple subsets of tweets to test on. Different sets of tweets have do not necessarily share the same vocabulary, which can be a signal itself, but does not guarantee consistent experiments in terms of word sets. As noted, not all of the target words and attribute words were present across all Twitter word embedding sets in our experiments, and we had to adapt accordingly. 

While we did not have validation data when we began our case study experiments, numerous reports have been released since then confirming Russia's involvement in spreading disinformation about COVID-19 on social media platforms, namely Twitter \cite{GECreport, NATOreport}. This confirmation of our results suggests that our method provides an effective method to identify information operations on emerging domains

\section{Conclusion}

Using a non-parametric AI method to quantify biases expressed on Twitter, our novel approach allows for real-time bias analysis of a given text corpus, without requiring expert annotated data. We adapt WEAT to measure bias associations for concepts central to information operations such as Calm/Panic and Trustworthy/Untrustworthy. Measuring these biases can help track how information operations spread chaos and distrust in targeted groups. We validate our method on Twitter data linked to known Russian and IRA information operations, selecting word sets that represent targeted information campaigns (\#BlackLivesMatter and the 2016 U.S. presidential election). We identify pro-Russian and anti-Chinese biases in recent COVID-19 related Twitter data. Various domains can apply this practical method by selecting the desired opposing targets (e.g., Russia vs. China) to discover and measure the present biases. This method could be used to characterize attitudes on social media platforms prior to major world events, such as the upcoming U.S. presidential election, or the quickly evolving COVID-19 outbreak, by automatically identifying emerging biases. If unexpected biases are detected, researchers might then examine whether these could be artificially and deliberately introduced to the public sphere.

\IEEEpeerreviewmaketitle

\bibliographystyle{IEEEtran}
\bibliography{references}

\end{document}